\documentclass[apj]{emulateapj}
\lefthead{TSUJIMOTO}
\righthead{A LACK OF MASSIVE STARS IN DWARF GALAXIES}

\def\ltsima{$\; \buildrel < \over \sim\;$}
\def\ltsim{\lower.5ex\hbox{\ltsima}}
\def\gtsima{$\; \buildrel > \over\sim \;$}
\def\gtsim{\lower.5ex\hbox{\gtsima}}
\def\ms{$M_{\odot}$ }
\def\msp{$M_{\odot}$}

\slugcomment{Accepted for publication in ApJ}

\begin{document}
\title{Chemical signature indicating a lack of massive stars in dwarf galaxies}

\author{Takuji Tsujimoto}

\affil{National Astronomical Observatory of Japan, Mitaka-shi,
Tokyo 181-8588, Japan; taku.tsujimoto@nao.ac.jp}

\begin{abstract}
Growing evidence supports an unusual elemental feature appearing in nearby dwarf galaxies, especially dwarf spheroidals (dSphs), indicating a key process of galaxy evolution that is different from that of the Galaxy. In addition to the well-known deficiency of $\alpha$-elements in dSphs, recent observations have clearly shown that $s$-process elements (Ba) are significantly enhanced relative to Fe, $\alpha$-, and $r$-process elements. This enhancement occurs in some dSphs as well as in the Large Magellanic Cloud, but is unseen in the Galaxy. Here we report that this feature is evidence of the lack of very massive stars (\gtsim  25\msp) as predicted in the low star formation rate environment, and we conclude that the unique elemental feature of dwarf galaxies including a low-$\alpha$/Fe ratio in some low-metallicity stars is, at least in some part, characterized by a different form of the initial mass function. We present a detailed model for the Fornax dSph galaxy and discuss its complex chemical enrichment history together with the nucleosynthesis site of the light $s$-process element Y.
\end{abstract}


\section{Introduction}

Nearby dwarf galaxies provide an excellent test bed for understanding how the Local Universe has evolved with cosmic time. This has been made feasible by 8-m-class telescopes, which can access  individual stars in nearby dwarf spheroidal (dSph) galaxies \citep{Tolstoy_09} as well as in the Magellanic Clouds \citep{Hill_04}. The detailed abundance data accumulated to date from a significant sample of stars enables us to chemically tag what occurred in these dwarf galaxies based on our knowledge of Galactic archaeology \citep[e.g.,][]{Freeman_02}.

A well-known chemical feature observed in dSphs is a lower $\alpha$/Fe ratio compared with that in the Galactic halo stars \citep[e.g.,][]{Shetrone_01,Venn_04}. Recent studies found the same level of [$\alpha$/Fe] in stars close to the lowest  metallicity \citep[e.g.,][]{Frebel_10a} and a knee in the [$\alpha$/Fe] vs.~[Fe/H] diagram \citep[e.g.,][]{Cohen_09}; these studies seem to support that the contribution of Fe from type Ia supernovae (SNe Ia), starting from a very low metallicity ([Fe/H] $\ll$-1) results in the observed low-[Fe/H], low-[$\alpha$/Fe] stars  \citep[e.g.,][]{Lanfranchi_03, Kirby_11}. However, no clear signature of an SN Ia enrichment for other elements, such as Mn/Fe and $n$-capture/Fe, casts doubt on this scheme \citep{Tsujimoto_06}. Moreover, a low-$\alpha$/Fe ratio has been detected for very low-metallicity stars \citep{Aoki_09}, and was also found in the Large Magellanic Cloud (LMC) \citep{Pompeia_08}. This observational fact seems to conflict with the presence of low-metallicity stars exhibiting high [$\alpha$/Fe] in some dSphs \citep{Koch_08, Venn_08, Frebel_10b, Letarte_10}, and clear answers to this problem continue to elude us.

In contrast, the abundance of $n$-capture elements, Ba and La,  are enhanced in some dSphs. First, we noticed that Ba (La) is enhanced relative to the $r$-process element, Eu, by examining an ensemble of data from six dSphs \citep[][references therein]{Venn_04}. Theoretically, the high [Ba/Eu] ratio can be explained with a strong galactic wind model, in which the $r$-process elements from type II SNe (SNe II) are no longer produced after the onset of the winds \citep{Lanfranchi_08}. This issue was remarkably furthered by the abundant data from a high resolution VLT study for the Fornax (Fnx) dSph \citep{Letarte_10}. This study revealed a clear increase in Ba and La, not only in comparison with Eu but also with Fe or  $\alpha$-elements in accordance with an increasing Fe/H during late evolution. This feature contrasts with that of the Galaxy and disagrees with current model predictions \citep{Lanfranchi_08}. A similar feature is seen in some other dSphs, for example in the Sagitarrius dSph \citep{Sbordone_07}, and clearly in the LMC \citep{Pompeia_08}. 

Based on an idea that invoking the initial mass function (IMF) variations may be one of the solutions to explain chemical abundance patterns \citep[e.g.,][]{Pagel_97}, we present a new mechanism underlying the observed trends between Ba, Eu, and Fe in some dwarf galaxies, with a focus on the Fnx dSph. Our claim is that star formations lacking very massive stars, such as $\gtsim$ 25 \msp, imprint an unusual $s$-process enhancement, which is  observed in stellar abundance. The suppression of the formation of massive stars in low surface-brightness galaxies is implied from the observed flux ratio of H$\alpha$ to the far ultraviolet \citep{Meurer_09}. In addition, from a theoretical aspect, recent work suggests that a high mass end $m_{\rm u}$ of the IMF depends on the mass of the star clusters, and thus on the physical properties of the galaxies where the clusters have formed \citep{Kroupa_03, Weidner_05, Pflamm_08}. These studies claim that $m_{\rm u}$ would be lower in the low density environment of dwarf galaxies, in which the formation of massive star clusters is suppressed. Their model predictions have been shown to be consistent with the observed trend for the H$\alpha$-to-FUV flux ratio \citep{Lee_09}. It is wothwhile to note that, in the Fnx dSph case, the approximate star formation rate (SFR) of 3$\times 10^{-3}$ \ms yr$^{-1}$ \citep{Coleman_08} is predicted to be $m_u$=$20-40$ \msp, while galaxies with a high SFR of \gtsim 1 \ms yr$^{-1}$ possess $m_u$=150 \ms \citep{Weidner_05}.

This proposed scenario must be reconciled with the framework based on the star formation history,  as revealed by extensive surveys for the Fnx dSph \citep{Battaglia_06, Coleman_08}. We show that this task leads to a grand scheme of chemical evolution for the Fnx dSph, in which a simple model is not applicable, but a complex view provides a deeper understanding of its elemental feature.

\section{Mechanism of an enhanced $s$-process abundance}

\subsection{Abundance patterns and nucleosynthesis}

A comparison of the Ba-Eu-Fe abundance trends between the Fnx dSph and the Galaxy is shown in Figure 1. Information on an early phase ([Fe/H]$\sim$-2.5) for the Fnx dSph is provided by the abundance of globular clusters (GCs) \citep{Letarte_06}, which must be equivalent to that of  interstellar matter at that time. On the upper panel, we see a sharply increasing [Ba/Fe] (roughly 0$\rightarrow$1) from [Fe/H]$\sim$-1 until [Fe/H]$\sim$-0.5. In contrast, in the Galaxy case this value is broadly constant at the level of [Ba/Fe]$\sim$0. On the other hand, for [Fe/H]$<$-1, the [Ba/Fe] ratio for the dSph remains nearly constant, and is on average higher than that of the Galaxy at [Fe/H]$\sim$-2.5. These two features of the Fnx dSph, i.e., a sharp increase at a late stage and an initially high ratio, can be understood from the theoretical aspect described in the following.

The key point to highlight is that the sites of nucleosynthesis among the $r$-, $s$-process elements and Fe are different from each other. Note that Ba is synthesized through both $r$- and $s$-processes. While Fe is synthesized and ejected in all SNe II with masses $\geq$10 \msp, it is believed that the site of $r$-process elements is confined to the limited SNe II that are inclined to be less massive. The theoretical interpretation of abundance data on very metal-poor stars implies that the mass range for the $r$-process is 8-10 \ms \citep{Mathews_92, Ishimaru_99} as identified with the collapsing O-Ne-Mg core \citep{Wheeler_98}, 20-25 \ms \citep{Tsujimoto_00},  or 12-30 \ms \citep{Cescutti_06} with heavy weighing on lower mass progenitors. On the other hand, the main $s$-process operates in AGB stars with a mass of 1.5-3 \ms \citep{Busso_01}, that might extend to the  more massive hot AGB ($\sim$ 6\msp) \citep[e.g.,][]{Goriely_04}.

\begin{figure}[t]
\vspace{0.2cm}
\begin{center}
\includegraphics[width=7cm,clip=true]{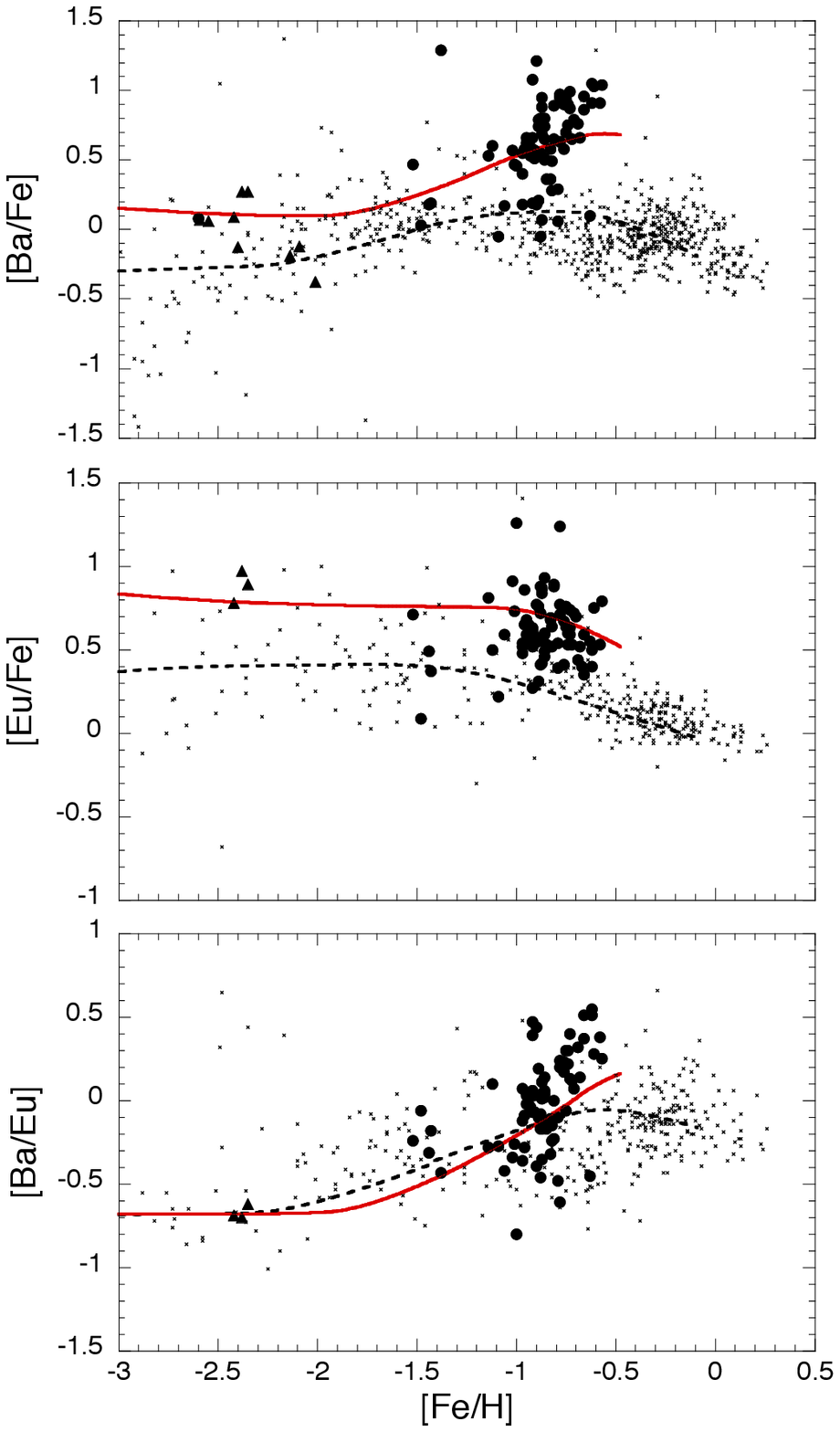}
\end{center}
\vspace{0.3cm}
\caption{Observed and predicted correlations of [Ba/Fe] ({\it upper}), [Eu/Fe]({\it middle}), and [Ba/Eu]({\it lower}) with [Fe/H] for the Fnx dSph, compared with those for the Galaxy. The observed data of the Fnx dSph are denoted by filled circles \citep{Letarte_10}, and those for globular clusters are denoted by filled triangles \citep[][]{Letarte_06}. The small crosses show the data for the Galaxy \citep{Venn_04}. The high-$s$ yield for Ba is assumed in the calculations. Two model curves are obtained for a different set of parameters in which an upper mass end of the IMF plays a principal role, set at 25 \ms (red curve: the Fnx dSph case -model Fnx1-) and 50 \ms (dashed curve: the Galaxy case), respectively. 
}
\end{figure}

Suppose that the Fnx dSph possesses the IMF with the high-mass end truncated around 25-30 \msp. Such a truncated IMF is motivated by a theoretical background \citep{Weidner_05} and also implied from the abundance of some metal-poor damped Lyman-$\alpha$ systems \citep{Tsujimoto_11}. The effect of its introduction is a reduction in the Fe mass ejected from each generation of SNe II, whereas the amounts of $r$- and $s$-process elements synthesized in low-mass SNe II and AGB stars, respectively, remain unchanged. As a result, for instance, the [$r$-process/Fe] ratio will become higher than that for the case with a normal IMF. Next, we examine the [Ba/Fe] ratio in the Fnx GCs at the metallicity of [Fe/H]$\sim$-2.5 (upper panel). From their low [Ba/Eu] ratios of $\sim$-0.7 (lower panel), Ba contained in the GC stars can be interpreted as a result of the pure $r$-process. Indeed, the high [$r$-process/Fe] ([Ba/Fe]$\sim$0) in the Fnx GCs,  compared with Galactic halo stars ($<$[Ba/Fe]$>$$\sim$-0.4) is realized. Consistently, high [Eu/Fe] ratios are seen in the same GCs as well (middle panel). Further  compelling evidence is found in [$r$-process/$\alpha$] in the Fnx dSph: our proposed IMF predicts a higher ratio for a whole metallicity range, which is confirmed by the observed high constant [Eu/$\alpha$] ratio ($\sim$+0.5) compared with the Galaxy case ($\sim$0) \citep{Letarte_10}. Similarly,  after an onset of the enrichment from AGB stars,  the $s$-process elements are more predominantly ejected than Fe. This process, together with a stagnant enrichment of Fe that results in the increased ejection of $s$-process elements in the same interval of $\Delta$[Fe/H] as in the Galaxy case, is expected to lead to a sharp increase in [Ba/Fe] (or [Ba/Eu]) against [Fe/H], as observed in the Fnx dSph.

\subsection{Model results}

To validate the above processes, we model two cases (the Galaxy and the Fnx dSph) to demonstrate how the different IMFs change the path of chemical evolution between the elements, Ba, Eu and Fe. The basis for the model is that the galaxy is formed through an infall of material from outside. With this framework, the SFR is assumed to be proportional to the gas fraction with a constant rate coefficient such as $\nu$=0.4 Gyr$^{-1}$ for the Galaxy disk \citep{Tsujimoto_10}. Here, $\nu$ is the fraction of the gas mass that is converted into stars per Gyr. For the infall rate, we apply a formula that is proportional to $t \exp(-t/\tau_{\rm in})$ with a timescale of infall of $\tau_{\rm in}$, which is assigned to $\tau_{\rm in}$=5 Gyr for the Galaxy disk \citep{Yoshii_96}. For the Galactic halo, the values of ($\nu$, $\tau_{\rm in}$)=(0.3, 0.1) are assigned for the duration of star formation $\Delta_{\rm SF}$=0.5 Gyr, thus reproducing the abundance distribution function (ADF) of halo stars \citep{Ryan_91}. The Galaxy IMF is assumed to be a power-law mass spectrum with a slope of -1.35, e.g., a Salpeter IMF, with a mass range ($m_l$, $m_u$)=(0.05 \msp, 50 \msp) \citep{Tsujimoto_97}. The IMF is always normalized to unity between $m_l$ and $m_u$, and is combined with the nucleosynthesis yields stated in the following paragraph as well as with the Fe yield of 0.64 \ms from SNe Ia \citep{Iwamoto_99}. It is assumed that the lifetime $t_{\rm Ia}$ of SN Ia progenitors spans over some range according to a distribution function $g(t_{\rm Ia})$.  Here we assume that the fraction $f_{\rm Ia}$ of the stars that eventually produce SNe Ia for $3-8$\ms in the solar neighborhood is 0.05 with a box-shaped $g(t_{\rm Ia})$ for $0.5\leq t_{ Ia}\leq3$ Gyr \citep{Yoshii_96}. Recent studies on the SN Ia rate in distant galaxies imply the different form of $g(t_{\rm Ia})$ favoring a large population of young progenitors for SNe Ia \citep{Mannucci_06, Sullivan_06, Totani_08, Maoz_10}. However, such a SN Ia frequency distribution is at odds with the presence of an observed plateau of [$\alpha$/Fe] among halo stars as well as thick disk stars. Thus, here we adopt the above form so that our model can reproduce the elemental features of the solar neighborhood stars.

Our study highlights the sites and yields of nucleosynthesis, and we consider them for three elements: Fe, Ba ($r$- and $s$-process), and Eu ($r$-process). First, our adopted Fe yield increases in accordance with an increase in the mass of the SN II progenitor for 10$<$$M<$50 \msp, which is implied both from the abundance pattern of very metal-poor stars and from an analysis of the SN light curve \citep{Shigeyama_98, Tominaga_07}. As the site for $r$-processes, we adopt the mass range of 20-25 \ms and the yields for Ba and Eu \citep{Tsujimoto_00}. We do not have to be exact in the mass range as long as it is set below $\sim$25 \msp. This is because its variance influences only the behavior of $r$-process enrichment at a very early epoch, outside the focus of this study. We consider two options for $s$-process yields, each of which provides  different timing for the onset of $s$-process enrichment. Observationally, the metallicity indicating this onset among the Galactic halo stars is difficult to detect. It is hidden by the large scatter in the abundance ratios, allowing two claims:  either an early start, such as at [Fe/H]$\sim$-2 \citep{Gilroy_88}, or a late start, at [Fe/H]$\sim$-1.4 \citep{Roederer_10}. Theoretically, the onset of $s$-process enrichment is mainly determined by the $s$-process yield synthesized in low-metallicity stars. If we adopt  the results of \citet{Busso_01} having mass range for $s$-process progenitors of 1.5-3 \msp, their very low-$s$ yield for low-metallicity AGB stars results in basically no contribution of the $s$-process to halo stars \citep[see][]{Cescutti_06}, and thus realizes a late start. We refer to this option as the low-$s$ yield case. On the other hand, to enable an early start, i.e., an onset of the influence of the $s$-process on very low-metallicity stars, we prepare another option for the $s$-process yield by diminishing the metallicity dependency for a low-metallicity regime in the above low-$s$ yield case. In this high-$s$ yield case, the wider mass range of 1.5-6 \msp,  including hot AGB stars, is adopted.

\begin{figure}[t]
\vspace{0.2cm}
\begin{center}
\includegraphics[width=7cm,clip=true]{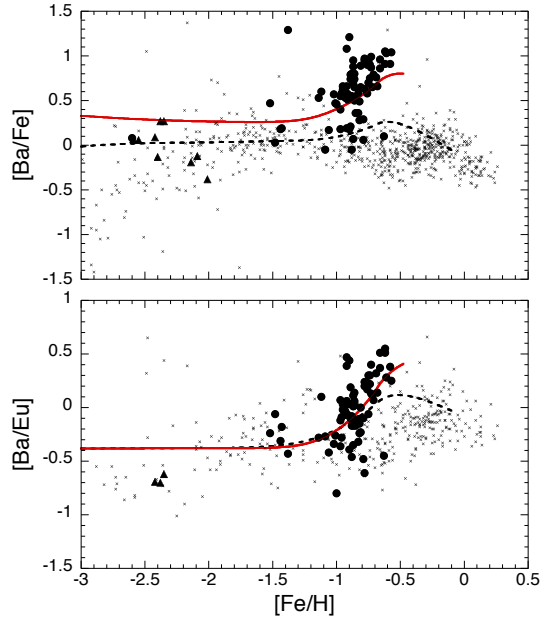}
\end{center}
\vspace{0.3cm}
\caption{The same as in Figure 1 except the model results are with a low-$s$ yield for Ba. 
The [Eu/Fe] vs. [Fe/H] plot is not shown since it is the same as the middle panel from Fig.~1.
}
\end{figure}

Figure 1 shows the results calculated with the high-$s$ yield, with the Galaxy models denoted by the dashed curves. Due to the high-$s$ yield, the [Ba/Fe] ratio gradually starts to increase from [Fe/H]$\sim$-2.2. After that the increase is suppressed due to a contribution of Fe from SNe Ia, and then the ratio decreases from [Fe/H]$\sim$-0.7, which is caused by a decreasing $s$-process Ba yield with the metallicity (upper panel).  A different trend is apparent in the evolution of [Eu/Fe], which is simply "a plateau plus a decreasing phase" in the same manner as the [$\alpha$/Fe] evolution. This trend is attributed to a negligible $s$-process production of Eu (middle panel). These two results lead to the [Ba/Eu] evolution as shown in the lower panel.

Next we examine the models for the Fnx  dSph, to determine if an increase in [Ba/Fe] at a late evolution can be reproduced. As discussed in the previous sections, this mechanism requires the introduction of a smaller $m_u$ for the IMF, set at $m_u$=25 \msp. In addition another setting is required for the model; that is, a relatively efficient star formation is demanded in order to prevent an increase in [Ba/Fe] from occurring at a lower metallicity than observed. This view is compatible with the observed ADF of stars that are older than 10 Gyr, which  extends to the metal-rich end around [Fe/H]=-0.5 \citep{Coleman_08}. Here, $\nu$=2 Gyr$^{-1}$ and $\tau_{\rm in}$=0.1 Gyr are chosen with $\Delta_{\rm SF}$=1.5 Gyr. The fraction of SNe Ia is determined to be $f_{\rm Ia}$=0.03 by the late evolution of Ba and Eu against Fe. We refer to this model as model Fnx1, since we will prepare for another two models in order to understand the chemical evolution of the Fnx dSph in \S 3. The results are indicated by the red curves in each panel. Each predicted curve passes through the GC data and broadly fits the observed trend at a late phase for -1.5\ltsim [Fe/H]\ltsim -0.5.

Results for the low-$s$ yield case are shown in Figure 2. This case requires the assumption of a  high [Ba/Fe] ratio ($\sim$0) as a plateau value for the halo, by raising the integrated $r$-process Ba yield by a factor of $\sim$2. According to this change, to suppress the increase in [Ba/Fe] at a very low metallicity for the Fnx dSph case, we reduce the $r$-process Ba yield by a factor of 0.6, compared to the $r$-process yield in the Galaxy case. These modifications are based on the hypothesis that the Ba yield for the mass $M\leq$25 \ms is broadly the same as in the high-$s$ yield case while the rest is synthesized in SNe II with $M>$25 \msp. Again we see good agreement of the model result assuming a small $m_u$ with the observed increasing Ba feature. 

\subsection{Insight into Y nucleosynthesis site}

Our study will be put forward to the examination into the nucleosynthesis of some other elements in the framework that connects their elemental features to our proposed view. Here we discuss the nucleosynthesis site of the light $s$-process element Y. Regarding $\alpha$-elements, detailed discussion will be presented in \S 4.

Yttrium ($A$=89) has been well observed in dSphs \citep[e.g.,][]{Shetrone_01}, including the Fnx dSph \citep{Letarte_10}. In regard to the $s$-process for the light $s$-elements with mass number 70\ltsim $A$\ltsim 90, the core He-burning in massive stars, the so-called weak $s$-process, is considered to be more important than the $s$-process operating in AGB stars. The weak $s$-process yield is found to increase with stellar mass \citep{Pumo_10}. Moreover, a recent study claims that the $r$-process for $A\sim$90-110 is produced by a charged particle reaction (CPR) and shares the same site with Fe production in massive stars \citep{Qian_07}. These theoretical results, together with the additional assumption of a similar mass-dependence between the $r$-process Y and Fe yields, provide a consistent interpretation of the Y feature present in the middle and lower panels. There is a broad coincidence of the [Y/Fe] value at the same [Fe/H] between the Galaxy and the Fnx dSph over an entire metallicity range, while there is a large deviation of [Ba/Y] between two galaxies in the late evolution. In Figure 3, the model results for the two galaxies are superimposed. Here we  adopt the Y yields which have a mass-dependence similar to Fe in massive stars for both $r$- and $s$- processes, and ignore the AGB contribution for Y.

\section{Chemical evolution of the Fornax dSph galaxy}

The model presented in \S 2.2 (i.e., model Fnx1) that reproduces the observed Ba feature is equipped with a rather short timescale (=1.5 Gyr) for star formation. Such rapid enrichment is observationally supported by the presence of old ($>$10 Gyr) stars having metallicity up to [Fe/H]$\sim$-0.5 \citep{Coleman_08}. On the other hand, the Fnx dSph has a much longer star formation history on the whole \citep{Battaglia_06, Coleman_08}. These two aspects seem to require the introduction of the complex model into the framework to consistently explain both the elemental feature and the overall star formation history. As a likely case, we can raise the possibility that different paths of  chemical enrichment exist leading to a large variation in the abundances of  the present-day Fnx dSh stars. It is, at least in part, inclined to result from the different speeds of star formation depending on the distance from the galaxy center, that will appear as an observed radial metallicity gradient \citep{Battaglia_06, Coleman_08}. 

\begin{figure}[t]
\vspace{0.2cm}
\begin{center}
\includegraphics[width=7cm,clip=true]{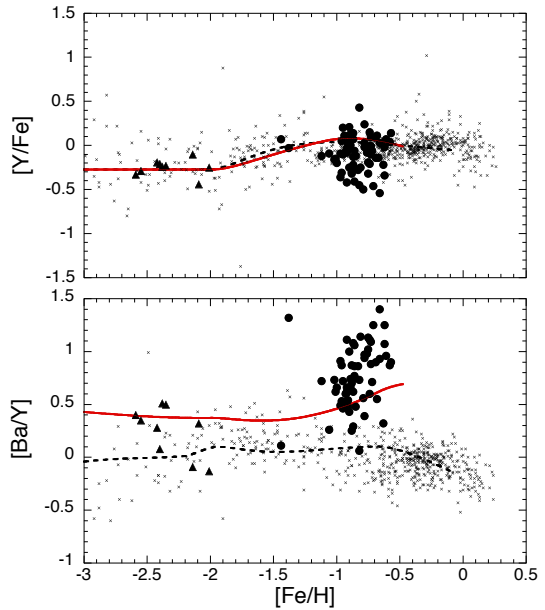}
\end{center}
\vspace{0.3cm}
\caption{Observed and predicted correlations of [Y/Fe] and  [Ba/Y] with [Fe/H] for the Fnx dSph (red curve), compared with those for the Galaxy (dashed curve).}
\end{figure}

In fact, a close look at the [Ba/Fe] evolution in Figure 1 shows that some stars in the Fnx dSph are located right on the predicted path for the Galaxy. This implied channel is modeled by ($\nu$, $\Delta_{\rm SF}$)=(0.08, 5) with a normal IMF, i.e., $m_u$=50 \msp. This case is referred to as model Fnx2. In addition, the presence of the age-metallicity relation in the Fnx dSph \citep{Battaglia_06, Coleman_08} implies that a large proportion of the metal-rich stars leading to an increasing Ba/Fe trend are relatively young. Incorporating the observed finding of an enhanced star formation at a very late epoch \citep{Coleman_08} into the model, we set ($\nu$, $\Delta_{\rm SF}$)=(1, 2) with $m_u$=25 \ms and an initial metallicity of [Fe/H]=-1.2 for this case, and refer to it as model Fnx3. Then, together with the first model presented in \S2.2, i.e., model Fnx1, the Ba/Fe evolutions predicted by the three models in total are shown in the upper panel of Figure 4. In the middle panel, the resultant ADF summed up by the individual models with a ratio of 0.5/0.3/0.2 is compared with the observation \citep{Battaglia_06}. The star formation history constructed by these relative contributions from individual models is compared with the observed history \citep{Coleman_08}, which is reassessed without counting very low star formation for $t<$ 2 Gyr and an age threshold of 13 Gyr, in the lower panel. 

\begin{figure}[t]
\vspace{0.2cm}
\begin{center}
\includegraphics[width=7cm,clip=true]{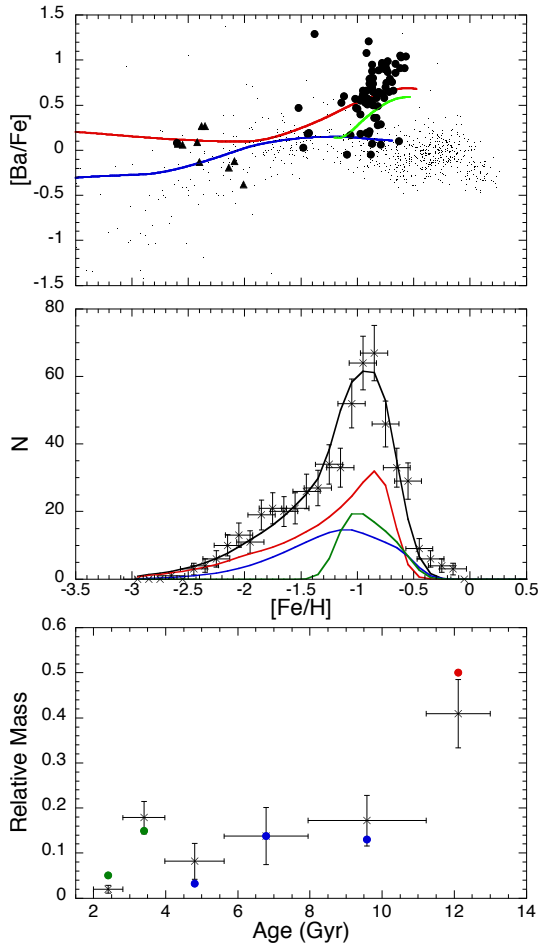}
\end{center}
\vspace{0.3cm}
\caption{Predicted chemical features of the Fnx dSph superimposed by three different paths of chemical enrichment, compared with observations. {\it Upper panel}: The [Ba/Fe] vs. [Fe/H]. Each curve with a different color corresponds to a different calculated path (red curve: model Fnx1, blue curve: model Fnx2, green curve: model Fnx3). {\it Middle panel}: Predicted ADF of the Fnx dSph stars against [Fe/H]. The calculated distribution is summed up by those by the individual model distributions, with a ratio of 0.5 (model Fnx1)/0.3 (model Fnx2)/0.2 (model Fnx3), and is convolved using the Gaussian with a dispersion of 0.1 dex considering a measurement error expected in the data. Each contribution by three models is shown by a different color. Crosses represent data taken from \citet{Battaglia_06}. {\it Lower panel}: Predicted relative stellar mass formed during each age bin (filled circles), compared with the observed results \citep[crosses:][]{Coleman_08}.}
\end{figure}

Here we review the adopted models in the context of the integrated galactic IMF (IGIMF) theory \citep{Kroupa_03, Weidner_05}. The IGIMF predicts that $m_u$ depends on the SFR (or gas mass) in a sense that $m_u$ is smaller according to a lower SFR. Thus, our hypothesis that model Fnx2 with a moderate SFR is equipped with a larger $m_u$ than models Fnx1\&3 with a higher SFR seems at odds with the IGIMF theory. As one possible explanation to integrate our models into this theoretical scheme, the stellar population associated with model Fnx2 may be regarded as that originally belonging to more massive building block within a hierarchical galaxy formation scenario. That casts our attention on the observed fact that the Fnx dSph obeys the luminosity-metallicity relation \citep{Kirby_08, Revaz_09}. This property may provide the possible path to the complex populations such that the merger event of Fnx2 into Fnx1 induces the enhanced star formation, and  yields a metal-rich component Fnx3, then ultimately ends up with a metallicity increase as a whole in accordance with the mass growth of galaxy. On the other hand, a maximal IGIMF model  together with the total mass of $1.9\times 10^7$\ms for the Fnx dSph \citep{Woo_08} approximately gives $m_u$=30, 20 \ms to model Fnx1, Fnx3, respectively. In this study, we have a close look at the signature of a small $m_u$ at an early evolutionary stage. However, given that we ignore it, one standard chemical evolution model that predicts a declining gas mass (SFR) as a function of elapsed time \citep{Kirby_11} is expected to lead to a gradual reduction in  $m_u$ that will make the Ba/Fe ratio mildly upward during late evolution.

\section{Understanding of the $\alpha$/Fe feature}

Finally, we examine the $\alpha$/Fe feature in the framework of the models constructed in the previous section. The prediction of the ratio of O (or Mg) to Fe ejected from an individual SN II still involves uncertainties because these two elements are synthesized by a completely different process. O (Mg) is produced through hydrostatic nuclear burning during stellar evolution before the SN explosion, whereas the amount of Fe is determined by the explosion mechanism and fallback dynamics, which are not fully understood yet. However, the observed feature of [$\alpha$/Fe] in dSphs cannot be fully interpreted by the knee structure positioned at a low-metallicity, since there exist stars exhibiting low [$\alpha$/Fe] ratios at a very low [Fe/H] \citep[see the distinct case of the Sextan dSph;][]{Aoki_09}. 

Suppose that the average $\alpha$/Fe ratio integrated over 10-25 \ms becomes small with the assumption that an increasing trend of yield against the progenitor mass is more prominent for $\alpha$-elements than for Fe. Here we assume that a truncated IMF in models Fnx1 and Fnx3 invokes that the average O, Mg, and Ti yields from SNe II are reduced by a factor of 0.7/0.5/0.3. Together with model Fnx 2, the results predicted by the three models for the Fnx dSph are shown by the colored curves, while the dashed curves represent the Milky Way case. For the nucleosynthesis yields of O and Mg form SNe II, we take their values tabulated in \citet{Tsujimoto_95}. On the other hand, the Ti yield as a function of progenitor's mass is deduced from the abundance analysis of very metal-poor stars as performed by \citet{Tsujimoto_98}, because the existing nucleosynthesis models fail to predict the Ti yield which is constrained form the observed level of [Ti/Fe] among halo stars \citep[see][]{Kobayashi_06}. For [O/Fe] and [Mg/Fe], the observed properties, i.e.,  a large scatter at a low-metallicity and an overall decreasing trend, are well reproduced. For [Ti/Fe], the observed data around [Ti/Fe] $\sim$+0.3 is lacking. However, medium-resolution observations suggest the presence of stars exhibiting such a high ratio \citep{Kirby_10}. Note that an unusual upward Ti/Fe feature for [Fe/H] \gtsim -1 is reproduced by the combination of low and high Ti/Fe yields in SNe II and SNe Ia, respectively. Since Ca is also synthesized highly in SNe Ia, our model predicts a similar upward feature for Ca/Fe. This is indeed observed in the Fnx dSph \citep{Letarte_10}. 

\begin{figure}[t]
\vspace{0.2cm}
\begin{center}
\includegraphics[width=7cm,clip=true]{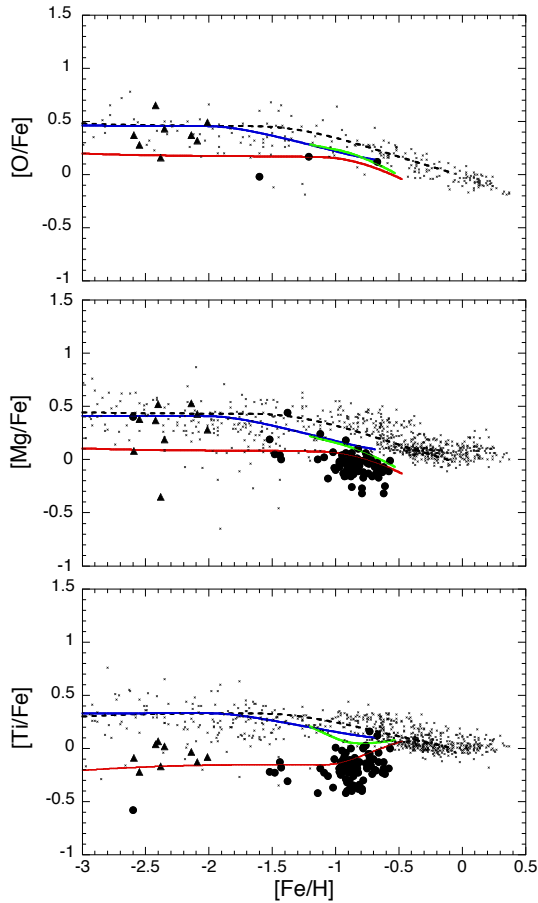}
\end{center}
\vspace{0.3cm}
\caption{Observed and predicted correlations of [O/Fe] ({\it upper}), [Mg/Fe]({\it middle}), and [Ti/Fe]({\it lower}) with [Fe/H] for the Fnx dSph, together with those for the Galaxy. Each curve with a different color corresponds to a different calculated path for the Fnx dSph (red curve: model Fnx1, blue curve: model Fnx2, green curve: model Fnx3),  while the dashed curve denotes the Galaxy case. For models Fnx1 and Fnx3, the modified nucleosynthesis SN II yields of $\alpha$-elements are adopted (see the text). Symbols are the same as in Figure 1 except for [O/Fe] \citep[filled circles:][crosses: Edvardsson et al. 1993; Gratton et al. 2000; Cayrel et al. 2004; Bensby et al. 2005]{Shetrone_03}.}
\end{figure}

Our model anticipates that the observed dispersion in both [Ba/Fe] and [Mg/Fe] at a low-metallicity is an end result of the IMF variation which yields a high [Ba/Fe] and a low [Mg/Fe] by $m_u$=25 \ms and a reverse correlation by $m_u$=50 \msp. Figure 6 demonstrates that the predicted correlation of [Ba/Fe] and [Mg/Fe] is broadly compatible with the observed one given by the data of the Fnx GC, though this analysis should be validated by much more data.
 
\begin{figure}[t]
\vspace{0.2cm}
\begin{center}
\includegraphics[width=7cm,clip=true]{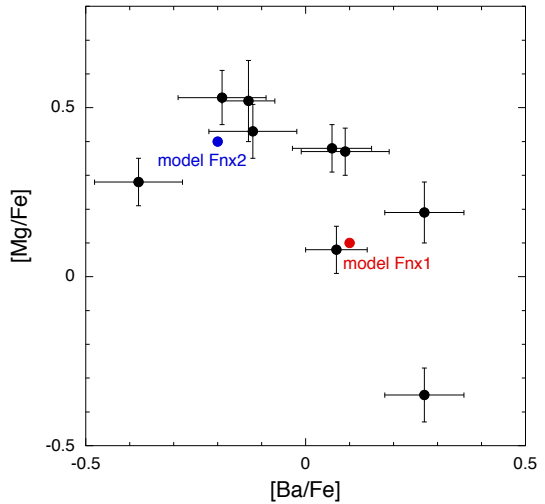}
\end{center}
\vspace{0.3cm}
\caption{The observed correlation of [Ba/Fe] and [Mg/Fe] for the GC in the Fnx dSph compared with those predicted by model Fnx1 ($m_u$=25 \msp) and model Fnx2  ($m_u$=50 \msp) at [Fe/H]$\approx$-2.5.}
\end{figure}

\section{Discussion and Conclusions}

Observed unusual Ba enhancement relative to Fe, $\alpha$-elements, and  Eu in some dSphs as well as in the LMC at their late evolution is investigated by modeling the Fornax dSph galaxy case. Our claim is that its effect occurs because the IMFs truncate a high mass end at around 25 \msp, that causes the reduction of $\alpha$-elements and Fe but no (little) influence on $r$- or $s$-process elements in  their ejection, associated with the death of stars covering a wide range of masses. Such a truncated IMF is assured by the theoretical agument given by the IGIMF scheme in which the number of massive stars depends on the SFR in galaxies. In addition, the star formation history of the Fnx dSph revealed by recent surveys, together with a large dispersion in elemental ratios such as [$\alpha$/Fe] and [Ba/Fe], suggests a rather complex chemical history not represented by a single model but composed of a few evolutionary paths resulting from a different speed of star formation as well as from a different form of the IMF. 

Previous work on the chemical evolution of dSphs interpret the lower [$\alpha$/Fe] ratios in these galaxies as being due to a combination of the time-delay model and a low SFR \citep[e.g.,][]{Carigi_02, Ikuta_02, Lanfranchi_03, Kirby_11}. This interpretation can explain the observed [Ba/Eu] ratios in dSphs with the inclusion of the galactic wind effect \citep{Lanfranchi_08}; however, it does not favor the same increasing [Ba/Fe] trend as [Ba/Eu] \citep[see, however,][as the model with a different assumption on galactic wind for the Sculptor dSph]{Fenner_06}. Further, it confronts the fact that a massive dwarf galaxy, i.e., the LMC, exhibits the same level of increase in [Ba/Fe] as would the much smaller Fnx dSph. On the other hand, our proposed model has the following shortcomings: (1) the lower [$\alpha$/Fe] is shifted to the uncertainties in theoretical nucleosynthesis in massive stars, (2) we lack an explanation for what the major driver of a complex star formation/chemical enrichment history is in the Fnx dSph, (3) we lack a sufficient explanation for why and how the high mass end of the IMF varies inside the Fnx dSph in the context of the IGIMF thoery. These issues point to the need for future studies.

\acknowledgements
The author wishes to thank an anonymous referee for his/her valuable comments and excellent review, that has considerably improved the paper, and is assisted in part by Grant-in-Aid for Scientific Research (21540246) of the Japanese Ministry of Education, Culture, Sports, Science and Technology.

\end{document}